\documentclass[12pt]{article}
\usepackage{graphicx}
\usepackage{amsfonts}
\usepackage{amsmath}
\usepackage{color}	
\usepackage{url}
\usepackage{authblk}
\usepackage[margin=0.5in]{geometry}

\title{Examining  collusion and voting biases between countries during the Eurovision Song Contest since 1957}
\author[1]{Alexander V. Mantzaris}
\author[1]{Samuel R. Rein}
\author[1]{Alexander D. Hopkins}
\affil[1]{University of Central Florida}

\begin{document}

\maketitle 

\begin{abstract}
  The Eurovision Song Contest (ESC) is an annual event which attracts millions of viewers.
  It is an interesting activity to examine since the participants of the competition represent a particular country's  musical performance that will be awarded a set of scores from other participating countries
  based upon a quality assessment of a performance. There is a question of whether the countries will vote exclusively according to the artistic merit of the song, or if the vote will be a public signal of national support for another country.
  Since the competition aims to bring people together, any consistent biases in the awarding of scores would defeat the purpose of the celebration of expression and this  has attracted researchers to investigate the supporting evidence for biases.
  This paper builds upon an approach which produces a set of random samples from an unbiased distribution of
  score allocation, and extends the methodology to use the full set of years of the competition's life span which has seen fundamental changes to the voting schemes adopted.
  By building up networks from statistically significant edge sets of vote allocations during a set of years, the results display a plausible network for the origins of the culture anchors for the preferences of the awarded votes.
  With 60 years of data, the results support the hypothesis of regional collusion and biases arising from  proximity, culture and other irrelevant factors in regards to the music which that alone is intended to affect the judgment of the contest.
\end{abstract}






\section{Introduction}
Our interconnected world brings together people from different countries and how they connect may depend on a miriad of factors. Efforts to understand social dynamics can be greatly assisted through statistical analysis of datasets gathered from communications and events which display preferences in connectivity.
In this paper the authors investigate the biases and collusion between countries participating the Eurovision Song Contest as a source of insight into the dynamics of different cultures joining together.

This contest is an annual event where countries have musicians present at a performance in which they will be given a score (for artistic merit) according to a voting rule set. At the end of the performances the song which received the largest number of points from other countries is declared the winner. With the purpose of bridging different nations together under a common shared interest in music which should transcend many barriers cultural, economic, language and history there has been a great deal of discussion whether these ties have been affecting the outcomes \cite{raykoff2007song,tragaki2013empire}.

The first of these competitions was in 1956 which is the only time countries were allowed to have more than a single song representing them and is therefore the only year this study ignores. There have been quite a few intricate regulations imposed on the performances such as  the number of singers allowed on stage, but many of these have changed and continue to do so. It is interesting to read more about the particulars, \cite{haan2005expert,clerides2006love}, especially as in 1957 there were only 10 countries mostly located around north west Europe, and the Eurovision Song Contest, now in 2017, has 42 participants (including Australia). The scale of the contest in terms of geography and the audience size puts the competition in the ranks of the largest reoccurring spectacles on the planet. 

Since some countries are going to be closer or further apart from each other there is likely to be some preferential voting according to geographic distances and the cultural distances. The paper of \cite{gatherer2006comparison} presented a versatile method and dataset which allowed readers to see the collusive voting behaviors between countries from 1975 to 2005. We present this approach of Gatherer in Section~\ref{subsec:algGatherer} explaining why it is a preferable approach to sampling from an unbiased distribution of score assignments. The  production of a distribution of unbiased scores allows us to examine the tails (in this case the upper tail) for a threshold of where we can then reject scores as originating from the same unbiased distribution but rather from a biased one when looking at the data. The Gatherer approach correctly uses multiple years rather than some other approaches making inferences from single point (estimates for each individual year), because the smoothing of a time window allows spurious activity to be have less weight in producing false positives which is important to avoid in high dimensional datasets \cite{abdi2007bonferroni,cox1962further}. The consistency of a behavior over a sequence of years is a more reliable way to examine whether there is statistical significance in the existence of a bias. Another reason that this approach is chosen to build upon is that it can be extended to accommodate different voting schemes with the extended framework developed here. The Table~\ref{table:scores} describes the dataset in more detail and the different voting schemes which were followed during the competition lifetime. In order to utilize the complete set of years in examining voting bias preferences, a flexible statistical framework is required. The underlying concept of generating an unbiased score distribution for each of the different voting schemes is presented in subsection\ref{subsec:extension}, which extends the Gatherer approach allowing us to choose sequences of years beginning in 1957 and ending with the most recent year 2017. 

This paper and that of  \cite{gatherer2006comparison} work with the definition of collusion as significantly large score assignments between two different countries over a set of years in comparison to a distribution of samples drawn from a hypothetical unbiased generator in the same conditions. Those results of  \cite{gatherer2006comparison} validate the hypothesis that certain pairs of countries (eg. Greece and Cyprus), and the Nordic countries (eg Sweden and Denmark) award each other with higher than expected scores. Since these behaviors have geographic and cultural overlaps it has led researchers to investigate this for the complete set of participants. Various computational modeling techniques have been used to answer whether there are political or cultural motives affecting the voting process, \cite{ginsburgh2008eurovision,budzinski2014culturally}. The work here helps answer questions such as these by providing a larger set of years to cross compare for stable and evolving relationships between different countries over a greater period of time and handling the changes to the award system.
This includes time windows that reference years  in which the voting schemes may not be homogeneous even within the chosen time span.

By examining the data for country pair sets showing significant indications of collusion (2-way pair voting bias) and one way bias (significant non-reciprocal voting from one country towards another) we can build a set of networks, Subsection~\ref{subsec:oneway} for collusion  and Subsection~\ref{subsec:oneway} for one way biases; which is not covered in \cite{gatherer2006comparison} but is considered in other research such as \cite{spierdijk2009structure,fenn2006does}.
The work of \cite{dekker2007eurovision} examines both collusive and one way biased voting links between countries and also emphasises the question of which countries act as 'bridges' between voting blocks which is something that the networks produced here help establish. 

The efforts to provide a complete analysis of the data gathered from the competition requires an understanding of the different voting schemes adopted throughout the lifetime of the competition (starting at 1957). The voting rules have changed a considerable number of times and a thorough investigation which can utilize all of the data must take this into account. In adopting a statistical methodology to produce the statistically significant set of biases from the data we look to ensure that the assumptions respect the data generating process which is not homogeneous in this case. Given the work of those already cited and including that of \cite{budzinski2014culturally,ginsburgh2008eurovision,saavedra2007identifying,blangiardo2014evidence,besson2007new} we see an active area of research which is likely to continue as the countries and contest evolves. A full literature review is not intended here, but the scope is to establish a new methodology which addresses the necessary question of how to fuse the timeline of voting histories produced from different voting schemes into a single consistent methodology that can provide answers of which participating countries displayed significant voting bias along any chosen set of years. As of yet a single unifying framework is not provided which can answer that question. Our work here provides in the public domain an easy to access dataset for the 60 years of scores from the selected judges and a methodology which allows us to investigate the countries which have awarded  significantly higher scores than should be expected without bias under any of the changing voting schemes adopted from 1957 till 2017.

\section{Data}

\label{sec:data}

The Eurovision Song Contest has undergone a substantial number of alterations in the
voting schemes it has applied since 1957. The need to introduce changes can be understood as the number of competing countries has been changing. This might induce a change in the distribution of the scores according to a varied set of factors and to some degree trial and error will be part of the organisation's goal to search for the best possible voting scheme. All of the data used is available at  \url{github.com/mantzaris/eurovision/}, which the authors aim to maintain up to date in the foreseeable years to come. For the convenience of the general audience the data of each year is placed in a separate csv table with intuitive headers. From the data sources the competition separates the votes that were chosen from a panel of judges and from televoting (general term for  residents who wish to express their choice for a winner from a country). Since the judge data is uninterrupted over the complete set of years, we take only these votes to provide consistency. It can be assumed that a panel of experts, who have a purpose of assigning scores purely on artistic credit, would not exhibit the behaviors we have set out to uncover such as collusion, but the regional clustering displayed in our results show otherwise. Having a bias within a group of judges only reinforces the concept that the country  specifics can distract the judgment even towards artistic expression when these different countries are brought together. 

The manner in which one country in the set of countries $\mathbf{c}$, gives votes to another country is denoted by $c_i$ where this quantity is the appropriate candidate number for these scores is $\mathbf{c}-1$ since a vote to their own country is prohibited.
The list of the different scoring schemes used in the contest since 1956 is shown in Table~\ref{table:scores}. Among the different methods, we can see 3 general paradigms for the score systems being; \emph{allocated/sequential/rated}. The \emph{allocated} label is given to the system where there is a fixed list of scores which a particular country can allocate one of the scores  towards another country's contribution. Each of the possible scores a country has to assign towards another country is allowed only once. Variations on how this scheme can be implemented is by having a change in the list of possible scores eg. 1962 and 1963 have a different list of scores to allocate ($[3,2,1]$ and $[5,4,3,2,1]$ respectively). The period of 2004 onwards has an 'allocated' scheme but it is important to note that there are rounds in which not every country participates in the final. (i.i.d. refers to independent and identically distributed)

With the \emph{sequential} voting scheme each country can receive one of the listed scores in sequence of the announcement. In terms of the mechanics of this procedure, it is described in more detail in the Subsection~\ref{subsec:extension}. Each country has an opportunity to receive one of the score labels each time where the score is drawn without replacement. 
The \emph{rated} scheme was tried only once for the years 1971-73, and had two jurors which would each give a score of $[5,4,3,2,1]$  towards each country participating. The sum of those two scores from each country would become the contestant's song award.
\begin{table}[h]
\begin{center}
  \begin{tabular}{|c|c c|}

  \hline
  \textbf{Years} & \textbf{Voting Scheme}       &  \\ \hline
  1956 & $[2]$ given to a single favourite       & \emph{(allocated)}\\  \hline
  1957-1961  & $\sum_{i=1}^{\mathbf{c}-1}c_i = 10$, each single point i.i.d. awarded       & \emph{(sequential)}  \\  \hline
  1962 & $[3,2,1]$ are mapped to 3 unique countries           & \emph{(allocated)}  \\  \hline
  1963 & $[5,4,3,2,1]$  are mapped to 5 unique countries       & \emph{(allocated)}   \\  \hline
  1964-1966 & $[5,3,1]$ given with possible consecutive awards            & \emph{(sequential)}   \\  \hline
  1967-1970 &  $\sum_{i=1}^{\mathbf{c}-1}c_i = 10$, each single point i.i.d. awarded            &   \emph{(sequential)}     \\  \hline
  1971-1973 & $c_i= X_1 + X_2$ each country $c_i$ gets a vote as a sum of 2 jurors $X_i \in [5,4,3,2,1]$        & \emph{(rated)}  \\  \hline
  1974 &  $\sum_{i=1}^{\mathbf{c}-1}c_i = 10$, each single point i.i.d. awarded           &      \emph{(sequential)}       \\  \hline
  1975-2003 & $[12,10,8,7,6,5,4,3,2,1]$ each country gives one of the scores to another country         &   \emph{(allocated)}    \\  \hline
  2004-2017 & $[12,10,8,7,6,5,4,3,2,1]$ the complete country set gives points to a final subset        &   \emph{(allocated)}   \\ \hline 
\end{tabular}

  \caption{Overview of the voting schemes used during each year of the Eurovision Song Contest and the label category each one is placed in.}
\label{table:scores}
\end{center}
\end{table}

\section{Methods}
\label{sec:methods}
Here we describe the algorithm of \cite{gatherer2006comparison}  for sampling from an unbiased model of score assignments between countries in Subsection~\ref{subsec:algGatherer}. Although the original paper outlines the operations of the algorithm, this Subsection does provide deeper insight into the general applicability of the methodology. 
 The extension proposed in Subsection~\ref{subsec:extension} provides a new methodology allowing the usage of the full lifetime of the competition, as the previous approach is compatible with only one of the voting schemes. The method develped here allows the analysis from a set of years where the scores were awarded according to more than a single voting scheme.

\subsection{Algorithm of Gatherer}
\label{subsec:algGatherer}
The operation of the algorithm for the calculation
of the null hypothesis that a country awards another country with scores
according to a uniform distribution is the  proposed method for
identifying statistically significant biases in \cite{gatherer2006comparison}.
In sampling the threshold for the score bias  between any two countries for
set of consecutive years the algorithm is presented in Table~\ref{table:Gatherer}. This algorithm calls upon another function presented in Table~\ref{table:votescheme1}.
\begin{table}[h]
\begin{enumerate}
\setlength\itemsep{-0.25em}
\item givenVotes = mean(data(country1,country2,startYear,endYear)  
\item simulationAverage = [ ]
\item for sample in sampleSize
\item \quad sampleScores = [ ] 
\item \quad for year in startYear to endYear:
\item \quad\quad participantsNum = participantsYears[year] 
\item \quad\quad position = ceil(rand(1,1) * (participantsNum-1))
\item \quad\quad if year $>=$ 1975 AND year $<=$ 2003
\item \quad\quad\quad score = allocationScheme(position)
\item \quad\quad\quad append(sampleScores,score)
\item \quad sampleAverage = mean(sampleScores)
\item \quad append(simulationAverage,sampleAverage)
\item sortedAverages = sort(simulationAverage)
\item threshold5percent = minimum(sortedAverages(1:(sampleSize*0.05)))
\item givenVotes $>$ threshold5percent ? print('collusion') : print('no collusion')
\end{enumerate}
\caption{The outline of the approach of Gatherer to produce a confidence interval for the threshold of the unbiased-biased score allocation between any pair of countries at the 95\% interval (line6 is described in Table~\ref{table:votescheme1}).}
\label{table:Gatherer}
\end{table}
Looking at the period 1975 till 2005 which the Gatherer paper examines, the voting scheme of \emph{allocation} applies. 
The score allocation is then defined according to \cite{fenn2006does}; where each voting country A allocates a set of points (1,2,3,4,5,6,7,8,10,12) to the ten other countries which are a subset of the entire set of S countries' minus one due to the fact that a country cannot vote for itself. By reversing the order of this score set, the position variable sampled in the above algorithm in line 9  is used to  produce the voting scheme in Table~\ref{table:votescheme1}.\\
\begin{table}[h]
  \begin{enumerate}
  \setlength\itemsep{-0.25em}
\item scores1975 = [12,10,8,7,6,5,4,3,2,1]
\item if position $<=$ 10
\item \quad score = scores1975[position]
\item else
\item \quad score = 0
\item end
\end{enumerate}
\caption{The outline of the \emph{allocation} voting scheme used in \cite{gatherer2006comparison} as a sampling scheme to produce a confidence interval for the threshold of the unbiased score allocation between any pair of countries for the 95\% interval}
\label{table:votescheme1}
\end{table}
To appreciate what this algorithm is achieving
for us we need to assess
what distribution is  effectively being sampled. In the situation where the countries participating ('participantsNum') is constant, the probability mass function is the multinomial distribution:
$P(\mathbf{x}) = P(x_1,\ldots,x_k) = \frac{N!}{x_1! \ldots x_k!}p_1^{x_1} \ldots p_1^{x_k} $, where $N$ is the number of years.
This can be mapped to the outcomes of a sequence of the contest results to produce a random allocation of results. This is effectively a null model in which we can compare against actual  votes  assess whether there is a large enough disparity in the average expected vote to support a statistically significant degree of  preferential voting.

The number of trials $N$ can be set to the number of years in \emph{startYear to endYear} in line 5, and each outcome $x_i$ corresponds to a particular score
a country can receive where $x_{i=1} = scores1975[1]$ and $x_{k=participantsNum}$ will be the lowest score available.
This score set of size $participantsNum$  is the number of countries participating where the
value can vary according to the rules during that year (line 6). The assumption is therefore that, $x_1=\ldots=x_k=0$. Sampling from this parameterization of the multinomial we could examine the size of the exponents after $N$ samples and map that to an aggregate score that is later sorted to produce a threshold for the 95 percent of cases produced with a value at least as large. There are multiple exponent permutations that can achieve this, eg. a score total over 2 consecutive years $N=2$ of some aggregate score of $y$  or more can be achieved with:
\begin{equation}
  \label{eq:multsum}
Score_{N=2}\left(x_1^2x_{2\ldots}^0, x_1^1x_2^1x_{3\ldots}^0, x_1^0x_2^2x_{3\ldots}^0,x_1^1x_2^0x_3^1x^0_{4\ldots}\right) >= y.
\end{equation}
The Gatherer approach used a window size of 5 years, and is equivalent to setting $N=5$ while performing an average of the samples.

In the introduction to this approach Gatherer mentions how the analytic formulation for the sampling is complicated and at times not possible due to the variation of the participant numbers (line 5). This can be achieved by sampling different distributions dependent on the  parameterization by the year and proceeding by then averaging as before. The comparison with the actual accumulated score average between pairs and this null model sample will still hold.
The expected values cannot be used since this would not allow us to reintroduce the score set by specific indexes. A change that we can choose is to eliminate the for-loop (line 4-9) and replace it by the loop required by the sequential trial set in the multinomial;
$j = \left( \sum\limits_{i=1}^{j'}x_i - u >= 0  \right)$ for each exponent index increment. This is done $1,\ldots,N$ where the samples $u$ are taken
from the uniform distribution $[0,1]$.

The multinomial approach in eq~\ref{eq:multsum} would require the mapping of each permutation to a score adding an overhead which the Gatherer algorithm does not have. From this we can see that the Gatherer approach is producing a statistically consistent sample set from the null distribution of scores in an efficient manner with prospects for accommodating when there are changes to the scoring scheme from participant numbers:
\begin{equation}
  P_{null}(score | participantsNum, endYear-startYear).
\end{equation}
When using this null model for taking samples it is equivalent to a Monte Carlo simulation by sampling the upon the hypothetical converged sample set of an equal outcome  upon the exponents of the multinomial for $N$.
This would be where the exponents of $x_i$, for $N$ samples, stating that 'over a set of $N$ contests one country gave another country each possible vote in equal proportion'. Sampling from the null model of $x_i^j= 1/k: \forall i \in [1,k]$ and the expected null sample of $j = c$; allows us to  examine whether if the average score given by one country to another during those years is greater than 95 percent of the samples drawn. Measuring convergence can be done by segmenting the sample set and looking at the stability of the mean score value, and one method is to compare the within and between variances. In our simulations a few thousand samples suffice for the analysis performed here, and given the reasonable sensitivity of the questions here using a rule of thumb that the confidence interval ceases to change in the third decimal place for a hundred samples should suffice as a stopping criteria.


\subsection{Algorithm extension for all voting schemes (1956-2017)}
\label{subsec:extension}
The Table~\ref{table:scores} describes the different voting schemes during the lifetime of the Eurovision Song Contest. We have 3 essentially different schemes which are best treated separately; allocated/sequential/rated.  The approach of Gatherer discussed in the original paper and in the previous Subsection~\ref{subsec:algGatherer}, is generalized here to handle the specific voting schemes for the years not included in 1975-2005 \cite{gatherer2006comparison}. Since each year is taken to be independent of the previous years, eg we assume that the score allocation a country has produced in a previous year does not have any effect on the subsequent years, and we can sample the uniform/unbiased score allocation for each year and examine the distribution of the total aggregate set. 

The periods for which the \emph{allocated} scheme applies is 1956/1962/1963/1975-2003/2004-2017 where modifications were made and are accounted for here. This can be accommodated for by changing the score list for each position randomly simulated in the algorithm. The change from 2004 onward incorporate rounds in which a subset of the countries can progress to a final and  only that subset of countries can receive votes and does not change the operation of the sampling when the correct participant number is taken into account for who can receive scores.  In general the assumption is that a score attributed towards a country is a uniformly sampled rank for each year. For each period of the allocated scheme the score in each position changes, eg in 1975 the fourth position receives 7 points but in 1962 it would receive zero. Thus, a country in each of these years allocates from a score set $s$ of length $|s|$ towards another country via an unbiased assignment, $x = \lfloor \emph{U}[1,c) \rfloor$:
\begin{equation}
\label{eq:allocated}
c_i =  \begin{cases} 
  s[x] & \text{if } x \leq |s| \\
   0  & \text{if } x  > |s|
  \end{cases}.
\end{equation}
This is outlined in Table~\ref{alg:allocated} which applies a simple case selection of the different allocation procedure for the score set that is directed towards candidate countries and the number of countries that participate.
\begin{table}[h]
\begin{enumerate}
  \setlength\itemsep{-0.25em}
\item function Sequential(yr,Num)\#year and country number
 \item  scores1 = [3,2,1]
 \item   scores2 = [5,4,3,2,1]
 \item   scores3 = [12,10,8,7,6,5,4,3,2,1]
 \item   position = ceil(rand(1,1)*Num)	
 \item   if(yr $>=$ 1975 \&\& yr $<=$ 2016)
 \item  \quad     scores = scores3
 \item   elseif(yr == 1962)
 \item  \quad     scores = scores1
 \item   elseif(yr == 1963)
 \item  \quad     scores = scores2
 \item   else
 \item  \quad     scores = scores3
 \item   end    
 \item   if position[1] $<=$ length(scores)
 \item  \quad    score = scores[position]	       
 \item   else
 \item  \quad    score = 0
 \item   end
\item return score
\end{enumerate}
\caption{The outline of the approach to simulating the unbiased \emph{allocated} voting scheme which applies to the years 1956, 1962, 1963, and 1975-2017.}
\label{alg:allocated}
\end{table}  
We are not considering the semifinal data which would allow us to monitor biases that might be missed (false negative), but since this part of the competition receives less public attention it is ignored. The year of 1956 is not included in the dataset, and in our study, because the only information that could be obtained was the outcomes and not the mappings of the awards between the countries participating.

The years 1964-66 had a voting scheme where a set of scores $s_1 = [5,3,1]$ was sequentially applied to the countries one or more times without replacement so that the total sum of the scores one country could allocate towards the candidates was $\sum_{j=1}^{3} s_1[j] = \sum_{i=1}^{c-1} c_i = 9$. This scheme permitted repetition and that had been observed. The years 1957-1961, 1967-1970, and 1974 used a similar scheme where single point is allocated i.i.d. across the candidate countries ten times. This allows repetition and in effect can be seen as another process of distributing a score vector's elements till the empty set is reached through draws, $\sum_{j=1}^{10} s_2[j] = \sum_{i=1}^{c-1}c_i = 10$ where $s_2 = \mathbf{1}$. This process of voting under the hypothetical situation of uniform unbiased awards over a set of years can be done by simulating a \emph{sequential} draw of each score and attributing it towards a country. This is presented in table\ref{alg:sequential} were it can be translated easily into actual code.
\begin{table}[h]
\begin{enumerate}
  \setlength\itemsep{-0.25em}
\item function Sequential(yr,Num)\#year and country number 
\item scores1 = [5,3,1]
\item scores2 = ones(Int,1,10)
\item score = 0
\item if(1964 $<=$ yr $<=$ 1966)
\item \quad  for ii=1:length(scores1)
\item \quad \quad  if(position = ceil(rand(1,1)*Num) == 1)
\item \quad \quad \quad score = scores1[ii] + score  
\item elseif(yr==1974 $||$ (1967$<=$yr$<=$1970) $||$ (1957$<=$yr$<=$1961))
\item \quad for ii=1:length(scores2)
\item \quad \quad  if(position = ceil(rand(1,1)*Num) == 1)
\item \quad \quad \quad score = scores2[ii] + score
\item return score
\end{enumerate}
\caption{The outline of the approach to simulating the unbiased \emph{sequential} voting scheme which applies to the years 1957-1961, 1964-1966, 1967-1970, and 1974.}
\label{alg:sequential}
\end{table}

The third voting scheme to be outlined for the unbiased sample generation was referred to as \emph{rated} in Table~\ref{alg:rated} for the years 1971-1973. This approach differs from the rest of the schemes in that the score vector $s_r = [5,4,3,2,1]$ is sampled from with replacement for each country twice and the sum of two independent samples is the awarded vote from one country to the next. If a sample from this set $s_r$ is denoted by $X$ the score for each country towards another one is independent of the number of country participants, $c_i = X_1 + X_2$. 
It was tried out for a short time and we can only speculate as to why it was never tried again. The effort to experiment with different schemes such as this one and be creative promotes the idea that the organizers are looking to avoid certain situations that arose in these three years. 
\begin{table}[h]
\begin{enumerate}
\setlength\itemsep{-.25em}
\item function Rated(yr,Num)\#year and country number
\item scores1 = [5,4,3,2,1]
\item    if(1971 $<=$ yr $<=$ 1973)
\item \quad  X1 = scores1[rand(1:end)]
\item \quad  X2 = scores1[rand(1:end)]
\item score = X1 + X2
\item return score
\end{enumerate}
\caption{The outline of the approach to simulating the unbiased \emph{rated} voting scheme which applies to the years 1971-1973. }
\label{alg:rated}
\end{table}

In a similar manner that the approach of Gatherer described in Table~\ref{table:Gatherer}, line 9 calls a function to produce a random sample of the unbiased sample of the allocated scheme, the approaches put forward here are combined by calling each of the Rated/Sequential/Allocated functions depending of the appropriate year to produce a consistent sequence of samples for a set of years.

\section{Results}
\label{sec:results}

As described in the methodology section we are looking at the voting history between countries for a chosen set of consecutive years to determine whether the scores exchanged are statistically significant indications of preferential voting. First examined are the two-way symmetric biases for a pair of countries which indicate collusion. Subsequently the one-way, as well as collusive pairs, are presented (subsection\ref{subsec:oneway}). A selection of time windows and years are chosen to provide a suitable demonstration without an exhaustive search of a particular configuration. This is done with the purpose to motivate other researchers to use the data, code, and methodology in exploring the support socio-economic theories which can be supported from this methodological application to the dataset.

The methodology builds upon that of \cite{gatherer2006comparison} to allow for the complete time line sampling along heterogeneous voting schemes. Other than the extension of the methodology used to account for years where the voting schemes differs from the years using the recent allocated scheme, two important features are altered. Although \cite{gatherer2007voting} does mention one-way biased edges, it is done without actually incorporating the investigation into the results which is given here giving further insight into the significant edges of score allocation. The other change is that within time windows if a country did not participate for the full set of years a new confidence interval as a threshold was calculated for the pair during the facilitating years producing lower values and increasing the number of collusive (two way biases) which is not done here. To produce a new threshold for the year subset is prone to false positives (erroneous edges) which is something chosen to be more important to avoid than the false negatives. Another feature that was not included are the semifinal results as the general population do not follow them as closely.

To group the countries we use a different approach to \cite{gatherer2006comparison}  choosing a similar style to that of \cite{spierdijk2009structure} which places them in (north, east, south, west).  Here the countries are organised into 6 different regions (north, north west, south west, central, south east, east), shown in Table~\ref{tab:regions} and the colors associated with these regions. Some of the smaller countries could be placed in different regions, eg Moldova being in the south east or east, and this ambiguity is reflected in the biases as is with some other countries. 

\begin{table}[!t]
 
\begin{center}
  \centering
	\begin{tabular}{ c | c  }
 	  southwest (red) & Portugal, Spain, Malta, SanMarino, Andorra, Monaco, Morocco, Italy\\ \hline
          northwest (turquoise) &  UnitedKingdom, Ireland, Belgium, France, Luxembourg \\
          \hline
	  north (blue) & Iceland, Denmark, Norway, Sweden, Finland  \\  \hline
          central (gray) & Germany, Austria, TheNetherlands, Switzerland, Slovenia, CzechRepublic,\\ & Hungary \\
          \hline
          southeast (orange) & Greece, Montenegro, Cyprus, Albania, Bulgaria, Croatia, BosniaHerzegovina, \\  & Turkey, FYRMacedonia Romania, Serbia, Israel, Yugoslavia \\ \hline
          east (green) & Russia, Ukraine, Moldova, Belarus, Poland, Georgia, Armenia, Azerbaijan, \\ & Estonia, Lithuania, Latvia
	\end{tabular}
	\caption{Regional label allocation of the countries and color attribute in the graphs}
	\label{tab:regions}
	\end{center}
\end{table}

\subsection{Collusion aggregates}
\label{subsec:collusion}
The collusion defined here is the presence of statistically significant voting bias between two countries over a set of years in a time window; and this is aggregated over the chosen period in the time windw. Presented in the images as a weighted network we can inspect the occurances. For example, if we look at a period of years 1975 to 1995 with year window 5, then there are four estimates of collusion between all the countries present in those incremental periods. Each pair of countries can then receive potentially four occurrences of significant collusion in those 5 year periods. We aggregate the count of collusion across these periods between each pair to examine the weighted networks.

\begin{figure}[!t]
\centering
a)\includegraphics[width=0.35\textwidth]{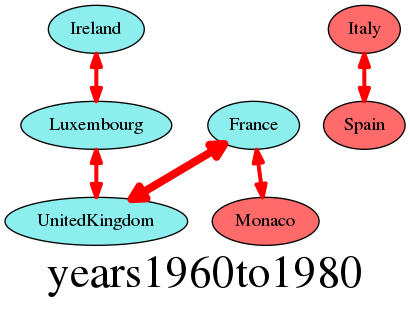}
b)\includegraphics[width=0.4\textwidth]{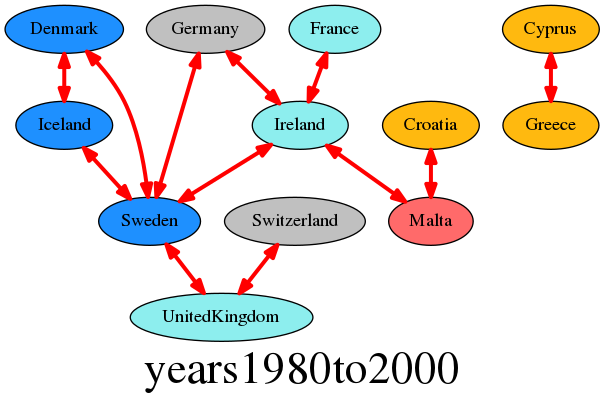}
\\
c)\includegraphics[width=0.99\textwidth]{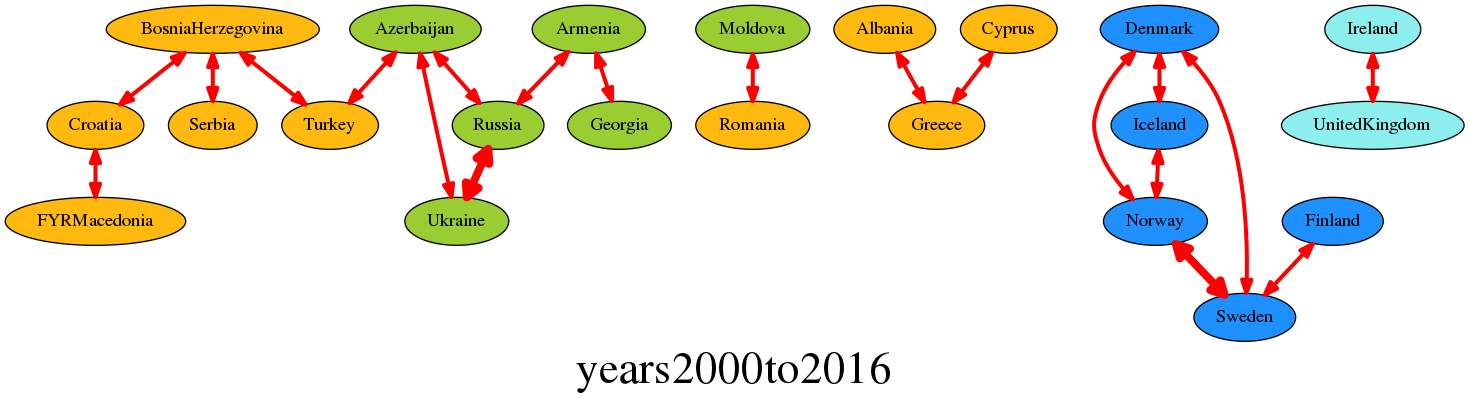}
\caption{Patterns of accumulated collusion for years 1960-1980, 1980-2000, 2000-2016 with time windows of 10, 10 and 8 years respectively. Thickness of the lines represents occurrence count.}
\label{fig:collusionSet1}
\end{figure}

In Figure~\ref{fig:collusionSet1} we examine 3 different time periods (1960-1980), (1980-2000), and (2000-2016) in the subfigures to measure the accumulated collusive edges between countries with time windows of 10, 10 and 8 years respectively. Subfigure a) shows that there is a clear link between the north western countries including Monaco. It would seem that Monaco could be included in the same set but when there are more countries we have seen a strong connection to Spain supporting a geographic similarity possibly due to culture of climate as can be seen in the Italy-Spain grouping when there are few other candidates in the set. Subfigure b) shows a clear image of the northern labeled countries with the United Kingdom, Ireland and Germany which are in their geographic neighborhood. It is quite common to see collusion between Greece and Cyprus which is well known and established but it is also interesting to point out the Croatia and Malta collusion. This pair appears in other examinations as well. Subfigure c) displays a great deal of geographic and cultural structure from the voting collusion although there are fewer years included. It attenuates the collusive patterns seen in other graphs as well. Given that we see the isolation of the northern countries establishes the bias over the previous years given that there was greater opportunity to produce another collaboration of score attribution. The south eastern countries form collusive blocks with interfaces to the eastern countries via those in closest geographic proximity (Turkey-Azerbaijan). This alludes to common economic/artistic/social features being shared for their to appear a collusion.



\begin{figure}[!t]
\centering
\includegraphics[width=1.0\textwidth]{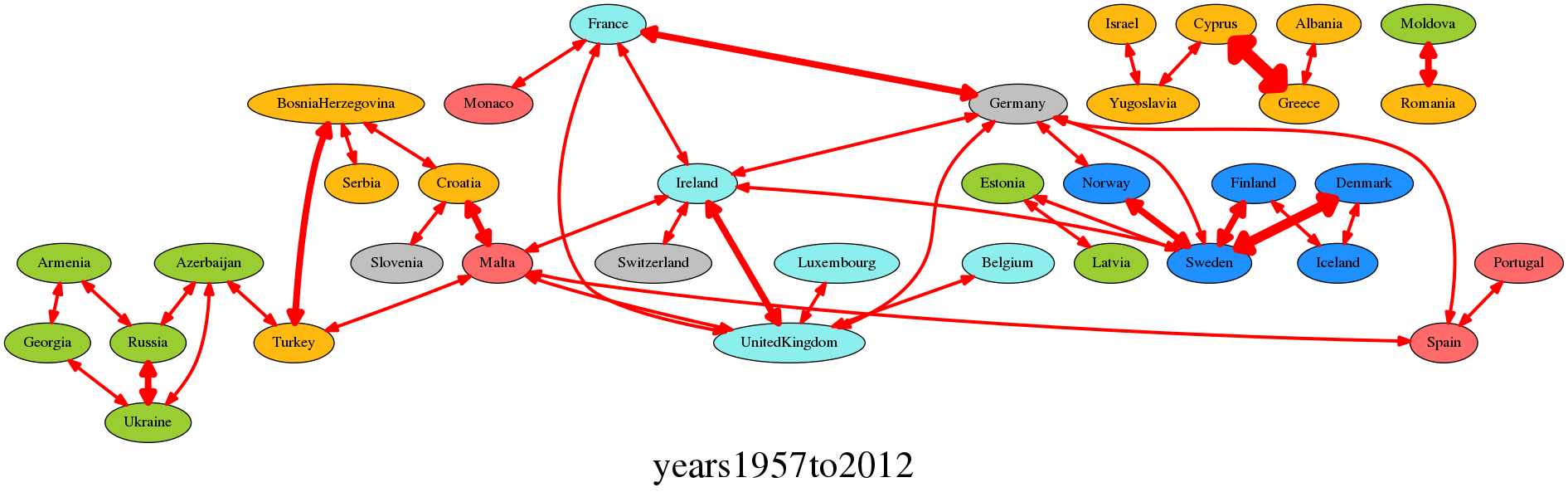}\\
\includegraphics[width=1.0\textwidth]{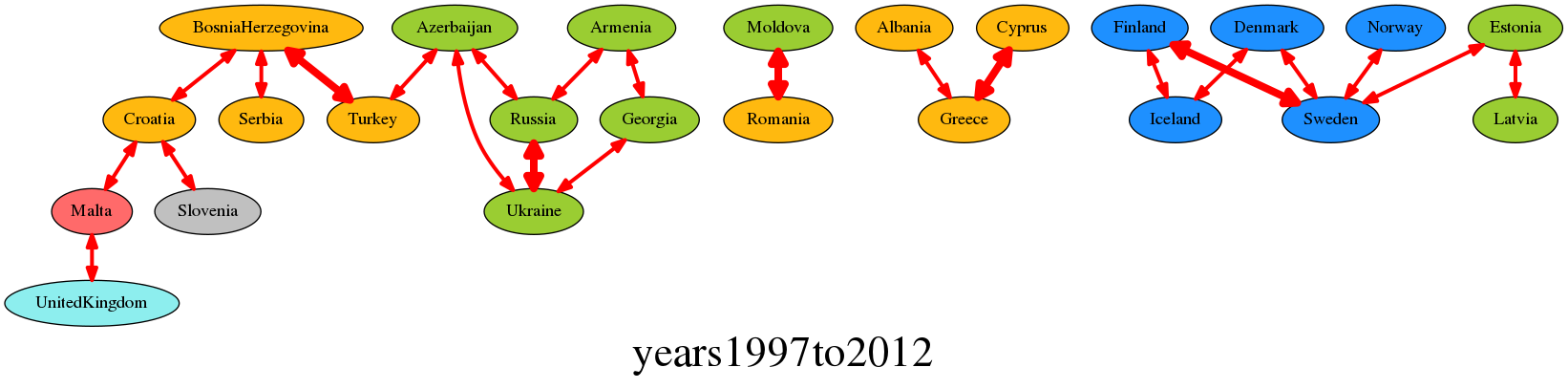}
\caption{Patterns of accumulated collusion for years 1957-2012 and  1997-2012 in subfigures a) and b) respectively with time windows of 5 years (thickness proportional to occurrences).}
\label{fig:collusionSet3}
\end{figure}

In Subfigure a) of Figure~\ref{fig:collusionSet3},  we see the years 1957-2012 covered in time windows of 5 years. It is easier to pick up on smaller stretches of collusion over the sequence of years and more edges can be added to the graph. We can clearly see the collusive tendencies in the geographic regions of the north, east, and south east. The south east's collusive behavior is split between themselves and their connections to the nearby eastern countries and Malta in the case of Croatia. It takes a moment to trace out some of the edges but the connectivity between the south western countries can also be seen with Spain, Portugal and Malta colluding. Ireland connects quite a few regions since in the early years of the competition there was a focus on its performances. This application samples over each of the 3 different voting schemes and variants using the methodology proposed here.


Subfigure b) of Figure~\ref{fig:collusionSet3} shows the years 1997-2012 with the time windows of 5 years which is comparable to that of Figure~\ref{fig:collusionSet3} with the major difference in that it excludes the early formative years of the competition. Removing the early stages of the collusive behaviors makes the network easier to see as the geographic pairing is now more clustered than what was previously.
It is almost evident in that Turkey acts as a bridge to the southern end of the eastern regions by connecting to Azerbaijan. Estonia bridges the east to the northern countries with collusion. It appears that the well known collusion between Greece and Cyprus has now brought in Albania showing a separate cultural or economic overlap that is emerging. Observations of changes such as this  can be used in conjunction with support from accomodating evidence that the clustering is dependent not only on geography but culture/language/economic ties as well when there are within region alterations. From the two subfigures we can see the effect of including more or less years while maintaining the same time window spacing differentiating between long standing collusive behaviors, transient ones and new associations. Figure~\ref{fig:collusionSet1} in comparison shows that with a larger window size the consistency within those competition years is more stringent to create significant collusive voting patterns.

\begin{figure}[!t]
\centering
a)\includegraphics[width=0.7\textwidth]{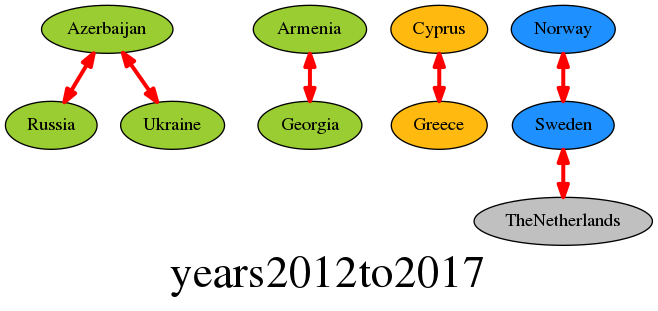} \\
b)\includegraphics[width=1.0\textwidth]{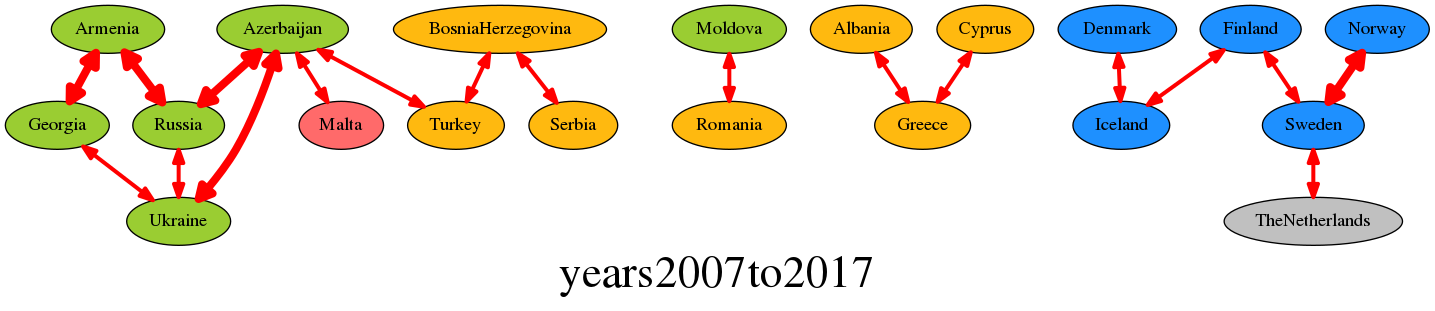}
\caption{Collusion for the periods 2012-2017, and 2007-2017 with a single year windows of 5 and then two 5 year windows respectively. Thickness of the lines represents occurrence count.}
\label{fig:collusionSet5}
\end{figure}

Figure~\ref{fig:collusionSet5} presents the collusion detected  in the past 5 years and then the previous two 5 year periods since 2007. We can see the persistent trend of the new eastern country collusive behaviors solidifying alongside the pre-existing ones of the northern countries and the south eastern countries beginning with Greece-Cyprus. It is important to note that the collusion between the south eastern countries being more numerous and dense appears to be producing a fraction not seen within the northern countries. The importance of certain countries in the southern parts of the east play an important role in connecting the different countries together.

We can see in the aggregate collusion networks a strong indication of geographic and cultural tendencies for positioning scores. The regional connectivity with cultural/language ties defines the collusive behaviors and it depends on the time periods to see this change. It is interesting to note that the original collusive behaviors are less dominant in recent years among larger countries or those with many participating neighbors. This can be explained by a dilution of biases not accumulating a majority in voting counties with larger absorbed groups and more neighbors. Overall it is possible to put forward the hypothesis that recent years of the competition show collusive behaviors increasing with distance from the central regions among countries with geographic proximity and cultural/economic ties. The number of biased edges that were reciprocated to produce collusion as a ration to those that were one way calculated for each of the plots shown was computed, and there is a trend that as the competition years are more recent the ratio of collusion to non-collusive score allocation is increasing. Which adds to the notion that countries are periodically increasing the locality to cooperate in a way to increase the chance of winning or produce public displays of support. 

The most important feature to note is the lack of edges that would appear to be spurious displays of random collusion. For example we do not see collusion between the United Kingdom and many small countries in the eastern or south eastern regions. That is explained by a lack of geographical proximity, cultural pollination,  or the ability for  a group from a smaller country forming a substantially large residency to affect the voting pattern. Many  examples like this exist when we fail to find edges such as Russia towards Iceland or between France and Ukraine. Figure~\ref{fig:collusionSet3} presents quite a few edges as it covers such a large time period but with the exception of the Ireland and Malta collusion, the edges follow this trend.

\subsection{One-way and two-way voting biases}
\label{subsec:oneway}
Here we look at statistically significant voting behaviors between countries allowing for both one-way or two-way significant biases to be accounted for. This  one-way directional indication of significant voting bias occurs when the number of accumulated points over the time window exceeds the threshold which can be expected under a uniform allocation of scores each year. The voting allocation can be reciprocated to produce a collusive pairing of votes awarded for the pair of countries as was presented in the Subsection~\ref{subsec:collusion}. When producing the set of edges, we distinguish these two situations as one directional and the other as collusive edges displayed as black single arrow edge and the collusive edge in a red bi-directional arrow.

\begin{figure}[!t]
\centering
a)\includegraphics[width=.35\textwidth]{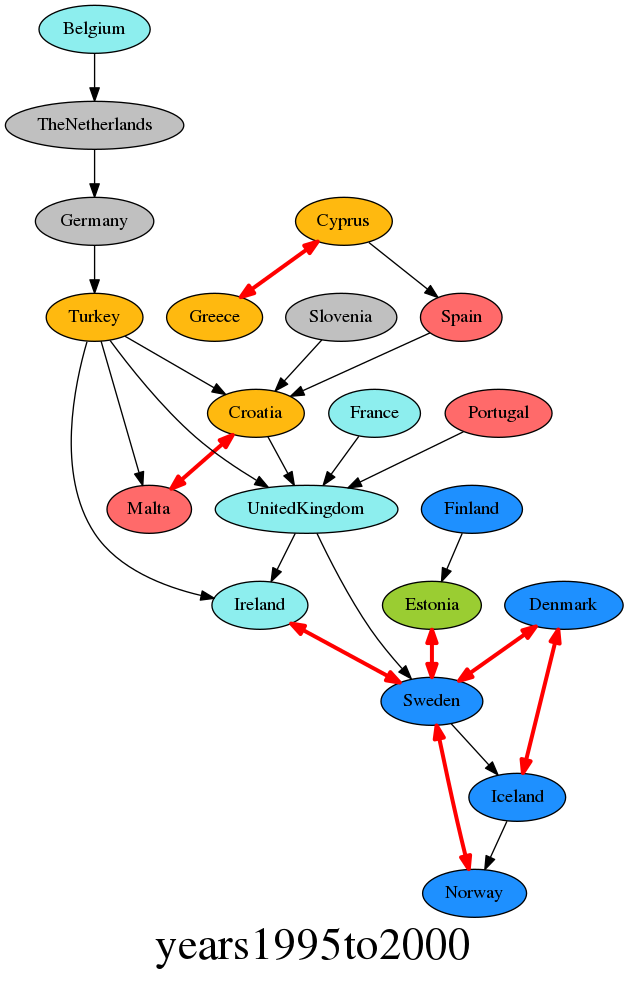}
\\
b)\includegraphics[width=0.75\textwidth]{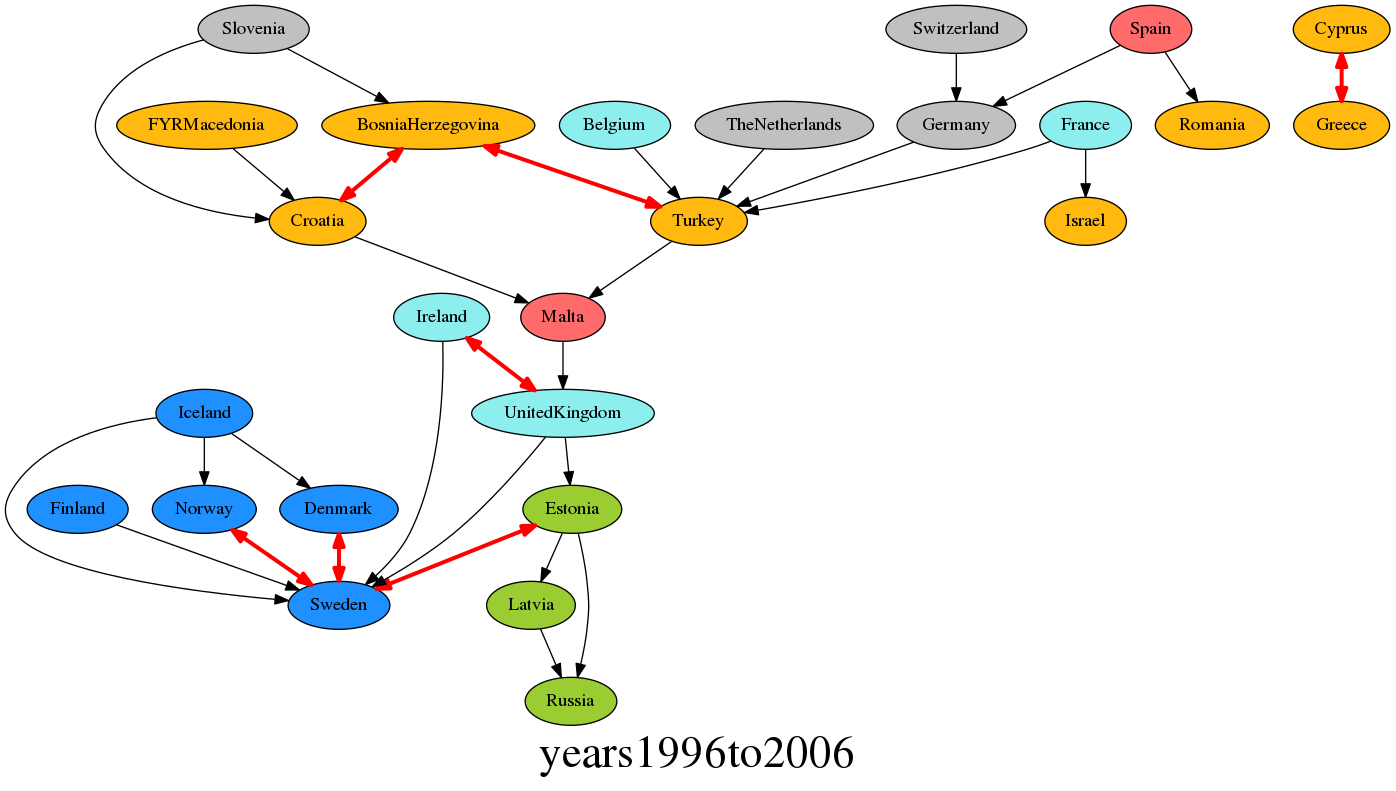}
\caption{Significant one-way and two-way (collusive) voting in the competition years  1995-2000 in subfigure a), and 1996-2006 in subfigure b) with a time window of 5 and 10 years respectively. The black directed arrows are for one-way biases and the red bi-directional arrows indicate two-way collusion.}
\label{fig:onewaySet1}
\end{figure}

In Figure~\ref{fig:onewaySet1} there are two subfigures for the years 1995-2000, and 1996-2006 each with a single time window 5 and 10 years respectively.
Subfigure a) encompasses two different applications of the 'sequential' voting scheme without requiring a separate analysis for each. As with Subsection~\ref{subsec:collusion} we can see that the collusive edges between the countries of the North region are present as well as the Greece-Cyprus pattern of collusion. 

Subfigure b) of Figure~\ref{fig:onewaySet1} shows the 10 year period of 1996-2006. With the introduction of the Eastern countries the bias between them can be seen as well as the established Northern regional cluster. The cluster also emerges with the South Eastern region as well. It is interesting how a small set of the countries provide the connectivity. In this graph Greece and Cyprus are disconnected and not in a) because the one-way bias between Cyprus and Spain was not consistent enough over the 10 year period in comparison with the smaller 5 year segment where the scores are concentrated for a plausible short term preferential voting bias. With this approach of window size alteration a consistency of the preferential bias can be examined.

The regional expression of bias is seen in the one-way as well as the collusive voting patterns shown in Subsection~\ref{subsec:collusion} with a large number of the cross regional biases being adjacent to each other. This is seen in the Estonia to Northern countries biases where cultural ties exist and many transit lines as well. The UnitedKingdom and Ireland one-way bias towards Sweden may have been overlooked if only the two-way collusions were taken into account. 
The connections like these and that of Greece-Cyprus as well as Sweden-Denmark show that two relatively low population groups with similar histories/languages/cultures separated by sea can display this behavior within the context of an event devoted towards artistic merit.

There are some interesting cross regional connections such as that of Germany to Turkey (in both subfigures) as well as Belgium/France/Netherlands  towards Turkey in b) which can be explained by the argument of \cite{haan2005expert}. Which puts forward the explanation that the migrant populations within these countries creates these observed connections, and that the lack of an equivalent migrant population would mean that this edge is not a symmetric collusive one. This feature would not be visible in the collusive network as it is not a symmetric bias that can be seen over multiple trials. This is another artifact of non-uniform voting which can have sociological importante. This puts value on the one-way significant edge detection beyond what can be considered a plausible false negative in a particular direction and a potential for further insight.
 It could be put forward that this is a mechanism of residing populations signaling their support for another country. This support direction appears to be focused towards adjacent countries and those from which there is a heritage connection.

\section{Discussion}

In this work, the authors have built upon previous research that has analyzed the Eurovision Song Contest (ESC) from 1975-2005 using a simple yet powerful approach to sample from a null hypothesis of uniform vote allocation between country pairs. A comparison of the historical vote assignment between countries can then be compared to this base case sample set where the value of  exceptionally large scores that falling outside of the range of confidence in uniformity are deemed to be derived from a biased set. With a comparative sample set it is possible to differentiate between  unbiased behavior of score allocation and biased score sets as significantly unlikely to have been produced under a uniform assignment. Reference models used to generate an unbiased uniform score set have been mostly applied to this period due to convenience as it is the longest time without a significant change in the voting schemes adopted. The improved methodology developed here offers allows  the inclusion of the various voting schemes introduced during the competition creation, and the ability to sample across time periods with heterogeneous voting schemes. With this addition, the examination of the contest can now be studied from its entire history (1957-2017) with any continuous time period in this range. 

A central point of research for the ESC, has been the detection of collusion, which is whether countries systematically award each other larger than expected scores over the years to provide an advantage in the competition or even if this is a display of political/cultural affiliation. Subsection~\ref{subsec:collusion} investigates the set of pairing  countries within a time period which exhibit significant preferential voting towards each other which form a graph. From the formative years of the competition the graphs show geographical proximity is an important factor for the collusive edges with clusters forming based upon regions. Many of the edges of collusion between different regions have a relatively low distance or plausible cultural affinity. Initially the strongest displays of collusion separated the north western, south western and northern countries before the contest included many of the south eastern and eastern countries. In the recent years, we see the continuation of the northern countries forming dense groupings as well as the eastern countries. The south eastern countries form clusters but commonly more than one cluster as their geographic region contains a high number of countries. A commonly mentioned collusive tie is the Greek-Cyprus which is shown here to exist in the analysis including data from the early years of the competition that now indicates a growth to include Albania. Many of the edges of collusion between central European countries do not appear consistent throughout the competition. In terms of regions the northern, eastern and south eastern countries exhibit most of the recent collusive behaviors.
It is interesting how various countries in a region act can as bridges between regions when they have ambiguous geographic placements as is the case for Azerbaijan in the east with Turkey in the south east. Other examples are Estonia with the northern countries and the eastern, Malta with the south eastern and north western.

The work here explores the ability to produce graphs built upon both collusive (two-way) and one-way edges in the same graph. The value of the one-way edges is that we can examine some of the biases not displayed in the collusive pairing graphs. These can arise through different situations such as a reciprocating edge being missing due to a dilution of a country's votes with geographically more neighbors in the vicinity or a resident population displaying solidarity from abroad to the country of their origin. 
The tendency for collusive edges to appear in their number among one-way biases is increasing steadily over time alluding that the countries find a benefit in reciprocating preferential treatment over the years to accumulate votes in an effort to win or see a need to display national support. 

It is also noted that the approach requires the user to choose a window size which can change the results. What is important is that regardless of the choice we see a consistent set of edges that support the hypothesis that regional proximity of the countries further from the central region display preferential voting behaviors towards their neighbors. The edges that span different regions that have larger distances are infrequently observed in collusion. A range of time periods is chosen for analysis reinforcing previous theories of preferential voting and regional bias.

The data used in this study is conveniently offered at \url{github.com/mantzaris/eurovision/dataTables} in the form of csv for each year will be kept up to date in the foreseeable future. The software used to perform this analysis was written in Julia \cite{bezanson2012julia,bezanson2017julia} which is a relatively new language offering many of the best features of other languages  in the area scientific computation (eg. Matlab, R, python). Running an analysis using the provided code requires only one line and three parameters of the years of choice and time window which is made clear in \url{github.com/mantzaris/eurovision/}. The networks are produced automatically with  \emph{graphiviz} and the authors encourage the readers to use the code to produce their own analysis.

Areas of future work are to analyze the effect on the collusion patterns based on the voting scheme adopted, and to associate the direction of the arrows with economic/social/cultural data between countries (such as trade agreements or linguistic differences).  The work of \cite{laufer2015mining} is an example where measures can be applied to quantify 'self-focus' and 'regional focus' in the context of cultural understanding to estimate the affinity, which can be combined with our conclusions. It would then be of interest to add  further comparisons with more subtle regional differences from competitions of national levels, as studied in \cite{budzinski2014culturally}, which would reinforce the concept of self-similarity being ubiquitous in all complex networks \cite{song2005self}.

\bibliographystyle{plain}
\bibliography{eurovisionRef} 


\end{document}